\documentclass[10pt,twocolumn,twoside]{IEEEtran}
\usepackage{graphicx,rotating,cite,subfigure,amsmath,array,amssymb,color}
\usepackage{url}
\usepackage{epsfig}

\newtheorem{theorem}{Theorem}

\newtheorem{lem}[theorem]{Lemma}

\newcommand{\qed}{\nolinebreak\hfill$\Box$\vspace{3pt}}
\newcommand{\pf}{\par \medskip \noindent {\bf Proof:} }

\def \cpha {{c}}
\def \lpha {{\ell}}
\def \sno  {{s}}
\def \rno  {{r}}
\def \dno  {{d}}
\def \sdl {{sd_\ell}}
\def \sdc {{sd_c}}
\def \sd {{sd}}
\def \sr {{sr}}
\def \rd {{rd}}

\allowdisplaybreaks[1]

\begin{document}
\bibliographystyle{IEEEtran}

\title{On Bounds and Algorithms for Frequency Synchronization for
  Collaborative Communication Systems}

\author{Peter~A.~Parker,~Patrick~Mitran,~Daniel~W.~Bliss,~and~Vahid~Tarokh%
  \thanks{This work was sponsored by the Department of the Air Force
    under Contract FA8721-05-C-0002.  Opinions, interpretations,
    conclusions and recommendations are those of the authors and are
    not necessarily endorsed by the United States Government.}%
  \thanks{Portions of this work are to appear in the Proceedings
    of the International Conference on Distributed Computing Systems,
    June, 2007.}}

\maketitle 

\begin{abstract}
Cooperative diversity systems are wireless communication systems
designed to exploit cooperation among users to mitigate the effects of
multipath fading.  In fairly general conditions, it has been shown
that these systems can achieve the diversity order of an equivalent
MISO channel and, if the node geometry permits, virtually the same
outage probability can be achieved as that of the equivalent MISO
channel for a wide range of applicable SNR.  However, much of the
prior analysis has been performed under the assumption of perfect
timing and frequency offset synchronization.  In this paper, we derive
the estimation bounds and associated maximum likelihood estimators for
frequency offset estimation in a cooperative communication system.  We
show the benefit of adaptively tuning the frequency of the relay node
in order to reduce estimation error at the destination.  We also
derive an efficient estimation algorithm, based on the correlation
sequence of the data, which has mean squared error close to the
Cram\'er-Rao Bound.
\end{abstract}


\IEEEpeerreviewmaketitle

\section{Introduction}

Collaborative communication systems employ cooperation among nodes in
a wireless network to increase data throughput and robustness to
signal fading.  Much of the research done in this area has
concentrated on information theoretic results, protocols, and coding
while assuming perfect synchronization~\cite{MitranOT05,
  SendonarisEA03i, SendonarisEA03ii, LanemanW03, LanemanTW04,
  JananiHHN04}.  In this paper, we explore frequency synchronization
of a collaborative system and provide estimation bounds and practical
algorithms having performance close to the bounds.

In a collaborative system, nodes that would have remained silent
during some period of time adapt to their surroundings and collaborate
with the source and destination nodes.  These systems, sometimes
termed cooperative diversity systems, use distributed protocols to
greatly improve performance over traditional point-to-point
communication systems. One improvement to system performance comes in
the form of added robustness to signal
fading~\cite{MitranOT05,SendonarisEA03i}.  An effective way to achieve
robustness is to increase the spatial diversity by using multiple
antennas as in a MIMO system~\cite{FoschiniG98,Telatar99}.  However,
when considering a network of low-cost wireless devices, the size and
cost of multiple antennas is prohibitive for these
devices~\cite{AkyildizSC02}.  A way for low cost nodes to realize much
of the benefit of a MIMO system is through collaborative (cooperative)
diversity.  In fact, in~\cite{MitranOT05} it is shown that a
collaborative system can have the same diversity order as an
equivalent MISO system.  Employing a collaborative protocol in a
wireless network can also increase the overall throughput of the
network.  The use of relaying is a special case of network coding and
as shown in~\cite{GastparV02}, the capacity of a relay (or coded)
network is greater than in a traditional point-to-point network.

To design a practical collaborative communication system, one of two
methods may be used. The signal modulation and coding may be designed
to be naturally robust to synchronization errors~\cite{Li04}, or
alternatively, the frequency and timing offsets are estimated and
subsequently compensated~\cite{ShinCKT05}.  We explore the second
option in this paper.  Algorithms and bounds for standard
synchronization are found in~\cite{BeekSB97, SchmidlC97, MorelliM00}.
The related case of a MIMO channel with multiple frequency offsets is
treated in~\cite{BessonS03, AhmedLJC05}.  In this paper, we provide
more details and extend the results of~\cite{ParkerMBT07}.  We derive
the transmission frequency the relay must use to optimally reduce the
variance of the frequency estimator at the destination by minimizing
the Cram\'er-Rao Bound~(CRB) of the frequency estimators at each
receive node.  By using the CRB, our frequency selection algorithm is
independent of algorithm choice.  We also provide an efficient
frequency estimation algorithm for the collaborative system.

\begin{figure}[t]
  \centering
  \includegraphics[width=3.2in]{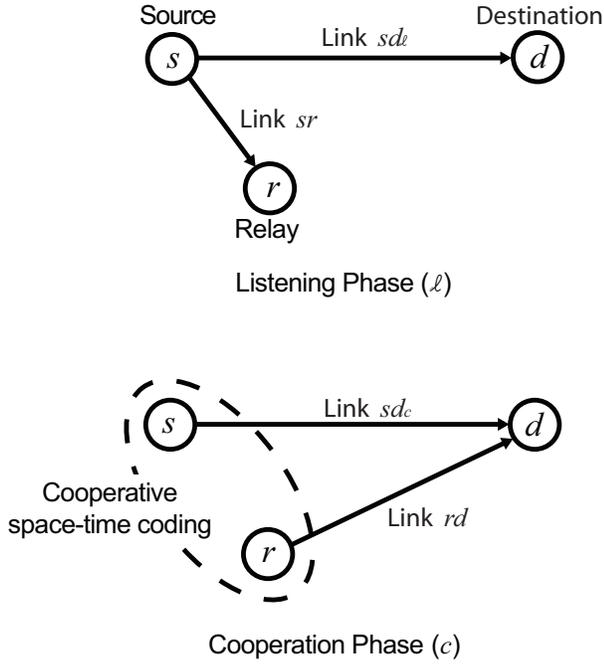}
  \caption{Illustration of the two phases in a three node cooperative
    communication system.}
  \label{fig:CoopSystem}
\end{figure}

In~\cite{ShinCKT05}, Shin et.~al.~describes a specific protocol, which
we use in this paper, for collaborative communication with
synchronization among three nodes: a source, a relay, and a
destination.  The protocol is based on a two-phase transmission within
each frame~\cite{MitranOT05,LanemanW03}, a listening phase and a
cooperation phase. Within each phase there is a preamble containing
synchronization signals.  In the listening phase, the relay receives
and decodes the source's message.  During the cooperation phase, the
relay re-encodes and transmits the message cooperatively with the
source.  This process is illustrated in Figure~\ref{fig:CoopSystem}.

The synchronization algorithms in~\cite{ShinCKT05} are ad-hoc and
meant only to serve as a proof-of-concept that synchronization is
possible with collaborative systems.  In this paper, we derive the CRB
for optimal frequency offset estimation for the class of systems
discussed above.  We show there exists an optimal (with respect to
minimizing the CRB) frequency of transmission for the relay node based
on: 1) the accuracy of estimation during the listening phase and 2)
the SNR of all node pairs.  We derive the maximum-likelihood~(ML)
frequency estimators for each receive node.  These estimators are
asymptotically efficient, meaning they achieve the CRB at high
signal-to-noise ratio~(SNR).  However, the ML solution is
computationally expensive and we therefore derive a practical
correlation based estimation algorithm with performance close to the
CRB.  For the purposes of this paper, we assume a frequency selective
fading model and that timing synchronization has been performed.
Future papers will extend this work to include timing estimation and
synchronization.  We also assume all training sequences are constant
modulus signals.

This paper is organized as follows, Section~\ref{sec:SysModel}
outlines the mathematical model describing the signals involved in the
frequency estimation portion of each phase.  The CRB and ML estimators
are derived in Sections~\ref{sec:Listen} and~\ref{sec:coop} for the
listening and cooperation phases respectively.  Section~\ref{sec:Sim}
provides some simulation results to illustrate the behavior and
performance of frequency estimation in the three node relay system
while Section~\ref{sec:Algs} shows the mean squared error~(MSE) performance
of each algorithm as compared with the CRB.

The following notation is used throughout: italic letters ($x$)
represents scalar quantities, bold lowercase letters ($\bf x$)
represent vectors, bold uppercase letters ($\bf A$) represent
matrices, $(\cdot)^T$ denotes transpose, $(\bar{\cdot})$ denotes
complex conjugation, $(\cdot)^H \triangleq (\bar{\cdot})^T$ denotes
complex conjugate transpose, $\|\cdot\|$ denotes the 2-norm of a
vector, $\Re(\cdot)$ denotes the real part of a complex number,
$\mathbb{E}_{\bf w}(\cdot)$ denotes the expectation operator with
respect to the random variable ${\bf w}$, $\mathcal{N}(\mu,\sigma^2)$
represents the Gaussian distribution with mean $\mu$ and variance
$\sigma^2$ and $\mathcal{CN}(\mu,\sigma^2)$ represents the circularly
symmetric complex Gaussian distribution, \emph{i.e.,} where the real
and imaginary parts are independent and identically distributed
Gaussian random variables with variance $\sigma^2/2$.

\section{System Model}
\label{sec:SysModel}

The system model is defined in this section.  During each phase, a
preamble consisting of a certain number of samples ($N_\lpha$ for
listening and $N_\cpha$ for cooperation) used for frequency
synchronization.  We assume the transmission channel is frequency
selective with channel impulse response $P$ samples long.  Due to
differences in local oscillator characteristics, the operating
frequency of each node is slightly different.  Let $f_\sno$ denote the
operating frequency of the source node and similar definitions for
$f_\rno$ and $f_\dno$.  The notation $\sd$ is used to denote the
source to destination link and likewise for $\sr$ and $\rd$.  As link
$\sd$ is used in each phase, let $\sdl$ denote the link during the
listening phase and $\sdc$ be for the cooperation phase.

Each transmitted signal is received and converted to baseband for
subsequent processing.  During the listening phase, the baseband
signal of link $a \in \{\sdl,\sr\}$ is expressed as~\cite{MorelliM00}
\begin{equation}
  \label{eq:SigMod1}
  y_a[n] = e^{j2\pi f_{a} n}s_a[n] + w_b[n], 
\end{equation}
where $n$ is the sample index, $f_a$ is the frequency offset between
the two nodes of link $a$ normalized by the sample rate, $w_b[n]$ is
the noise generated in the electronics of receiver $b\in
\{\dno,\rno\}$ (destination or relay node respectively), and $s_a[n]$
is the combination of the known training signals (${\bf x}_\lpha =
[x_\lpha[0], \ \dots, \ x_\lpha[N_\lpha-1] ]^T$) and the effects of
the frequency selective channel, given by
\begin{equation*}
s_a[h]=\sum_{k=0}^{P-1}h_a[k]x_\lpha[n-k].
\end{equation*}
In this equation, $h_a[n]$ are the samples of the channel response
for link $a$ and $P$ is the duration of the channel response.  We
assume, for each link $a$, the length of the channel $P$ is
the same.  Writing~\eqref{eq:SigMod1} in matrix form gives
\begin{equation}
  \label{eq:ListenSig}
  {\bf y}_a = {\bf V}_{f_a}{\bf X}_\lpha{\bf h}_a  + {\bf w}_b
\end{equation}
where $[{\bf V}_{f_a}]_{n,n} = e^{j2\pi f_a n}$ is a diagonal matrix
and $[{\bf X}_\lpha]_{i,k} = x_\lpha[i-k]$ is a Toeplitz matrix where
$x_\lpha [k] = 0$ for $k<0$ and $k\geq N_\lpha$.

In the cooperation phase, the signal is defined as follows,
\begin{equation}
  \label{eq:CoopSig}
  {\bf y}_\cpha = {\bf V}_{f_{\sd}}{\bf X}_{\sdc}{\bf h}_{\sdc}
  +{\bf V}_{f_{\rd}}{\bf X}_{\rd}{\bf h}_{\rd} + {\bf w}_\dno,
\end{equation}
where we assume the frequency $f_{\rd}$ is constant over both
phases. For each receiver, $b$, the noise is assumed to be a zero-mean
circularly symmetric complex Gaussian random vector
\begin{equation*}
  {\bf w}_b \sim \mathcal{CN}({\bf 0},\sigma^2_{b}{\bf I}).
\end{equation*}

In the general case, the frequency offsets between nodes can take on
any values within the Doppler spread of the system plus the frequency
differences of the local oscillators.  We assume the maximum frequency
offset is bounded and use this information to calculated the CRB and
ML frequency estimators.  In the remainder of the paper, we assume the
nodes are stationary and thus the signals have no Doppler spread.  A
statistical model for the frequency offset is used as prior
information to aid in frequency estimation.  Let the operating
frequency of each node $m\in\{\rno,\sno,\dno\}$ be modeled as
\begin{equation*}
  f_m = f_o + q_m,
\end{equation*}
where $f_o$ is the mean operating frequency and $q_m$ is a random
variable with mean zero and variance $\sigma^2_m$.  We assume the
random variables $q_m$ are independent. For this paper, we also assume
$\sigma_m^2 = \sigma_f^2$ for all nodes $m$, which is an appropriate
model when considering a group of identical nodes cooperating
together.  The frequency offsets to be estimated are the difference
between two of these independent random variables and thus the
frequencies, $f_a$ for $a \in \{\sd,\sr,\rd\}$, have mean zero,
variance $2\sigma_f^2$, and are correlated.

\section{Listening Phase}
\label{sec:Listen}

In the listening phase, the destination and the relay receive the same
signal through two different channels. We drop the subscript $a$ when
considering only the single node-to-node link.  To derive a good
estimator for the frequency, it is useful to know the distribution of
$q_m$.  However, this is not known, so it is reasonable to design an
estimator based on the ``worst case'' distribution constrained to the
known statistics, \emph{i.e.,} a mini-max estimator.  As frequency
estimation is inherently non-linear, an asymptotic analysis is
performed in the high SNR regime (\emph{i.e.,} SNR$\gg1$).  Under this
assumption, the variance of a ML or maximum a posteriori~(MAP)
estimator is equal to the CRB.  In the remainder of this section, we
show that a Gaussian distribution with mean zero and variance
$\sigma_f^2$ for $q_m$ maximizes the CRB of the frequency estimate
over all distributions with the same mean and variance.  We then
derive the MAP estimator of $f$.

\subsection{Cram\'er-Rao Bound}

The unknown parameters in the single node-pair
model~\eqref{eq:SigMod1} are $f$ (which is modeled as a random
variable with mean zero and variance $2\sigma_f^2$) and ${\bf
  h}$.\footnote{The parameter $\sigma_f$ is considered known as it is
  a property of the receiver hardware. Also, the noise variance
  $\sigma^2$ is uncoupled with the other parameters and is estimated
  separately with no penalty.}  The CRB is defined to be the diagonal
entries of the inverse Fisher Information Matrix~(FIM). When one or
more parameters are random variables, the FIM is expressed in the
following form~\cite{VanTrees1}
\begin{equation}
  \label{eq:StochFIM}
  {\bf J}_{\bf \theta} = \mathbb{E}_f({\bf J}_{\theta|f}) + {\bf J}_{f},
\end{equation}
where the expectation is taken over the random variable $f$,
\begin{equation*}
{\bf J}_{\theta|f} = -\mathbb{E}_{\bf w}\left(
\frac{\partial^2}{\partial \theta \partial \theta^T} L({\bf y}|f)
\right)
\end{equation*}
is the standard (non-random parameter) FIM, with expectation over the
noise distribution, and $L({\bf y}|,f) \propto
\frac{-1}{\sigma^2}\|{\bf y} - {\bf V}_f{\bf Xh}\|^2$ is the
log-likelihood of the data vector when the values of ${\bf h}$ and $f$
are held constant.  The matrix ${\bf J}_f$ is defined as follows:
\begin{equation*}
  {\bf J}_{f} = -\mathbb{E}_{\bf f}\left( \frac{\partial^2}{\partial
    \theta \partial \theta^T} L(f) \right),
\end{equation*}
where $L(f) = \log p(f)$ and $p(f)$ is the distribution function of
the random variable $f$.
For the parameter vector $\theta^T
= [f \ {\bf h}^T \ \bar{\bf h}^T]$, the FIM has the following
form~\cite{Smith05}
\begin{equation}
  \label{eq:FIMStructure}
  {\bf J}_{\theta|f} =
  \left[\begin{array}{ccc} {\bf \Delta} & {\bf \Lambda} &
      \bar{\bf \Lambda}\\ {\bf \Lambda}^T & \mathbf{0} &
      {\bf \Xi}^T\\ \bar{\bf \Lambda}^T & {\bf \Xi} &
      \mathbf{0} \end{array}\right],
\end{equation}
where ${\bf \Delta}$ is a scalar in this case.  Let $[{\bf D}_\lpha]_{nn} =
2n - 1 - N_\lpha$ be a diagonal matrix such that
\begin{equation}
 \label{eq:DefD}
\frac{\partial}{\partial f}{\bf V}_f = j\pi{\bf D}_\lpha{\bf V}_f.  
\end{equation}
The submatrices of~\eqref{eq:FIMStructure} are computed as
\begin{align}
  {\bf \Delta} &= \frac{2\pi^2}{\sigma^2}\|{\bf D}_\lpha{\bf Xh}\|^2\nonumber\\
  {\bf \Lambda}& =  
  \frac{-j\pi}{\sigma^2}{\bf h}^*{\bf X}^*{\bf D}_\lpha{\bf X}\nonumber\\
  {\bf \Xi}& = \frac{1}{\sigma^2}{\bf X}^*{\bf X}\nonumber.
\end{align}
None of these components depend on the random variable $f$ and
therefore the expectation in~\eqref{eq:StochFIM} goes away.  The
matrix ${\bf J}_f$ is only non-zero in the first element and is
\begin{equation}
  \label{eq:FIf}
  [{\bf J}_f]_{11} =
  -\mathbb{E}\left(\frac{\partial^2 L(f)}{\partial f^2}\right)
 \triangleq  F_f,
\end{equation}
where $F_f$ is the Fisher information of the random variable $f$ and
$L(f)$ is the log-likelihood of $f$.  The CRB for an estimator of $f$
is then $[{\bf J}_{\bf \theta}^{-1}]_{11}$, which can be calculated
using the Shur complement\footnote{The $\{1,1\}$ block of a block
  matrix inverse is $[A^{-1}]_{11} = (A_{11} - A_{12}A_{22}^{-1}A_{21}
  )^{-1}$.}~\cite{MoonStirling} to be
\begin{equation*}
  C_f = \left(\frac{2\pi^2}{\sigma^2}
    \| \mathbb{P}_{\bf X}^{\perp}{\bf D}_\lpha{\bf Xh}\|^2 +
F_f \right)^{-1},
\end{equation*}
where $\mathbb{P}_{\bf X}^{\perp} = {\bf I} - {\bf X}({\bf X}^*{\bf
  X})^{-1}{\bf X}^*$ is the projection matrix onto the space
orthogonal to the range of ${\bf X}$.  As the Fisher information is a
positive number, it is clear that, to find the worst case (maximum)
CRB, $F_f$ must be minimized.  We use the following Lemma to show how
this variable is minimized.
\begin{lem}
  \label{lem:MinFI}
  Let $p_{\sigma}(\cdot)$ represent the family of distributions with
  mean zero and variance $\sigma^2$.  Let $z$ be a random variable
  distributed as $p_{\sigma}(z)$.  The minimum of the Fisher
  information of $z$, as defined in~\eqref{eq:FIf}, over the family of
  distributions with variance $\sigma^2$ is achieved when
  \begin{equation*}
    p_\sigma(z) = \mathcal{N}(0,\sigma^2).
  \end{equation*}
  \pf Consider the following experiment: without any data, design an
  estimator $\hat{z}$ for the random variable $z$.  The log-likelihood
  in this case is $L(z) = \log p_\sigma(z)$. If $\hat{z} = 0$, then
  this estimator is unbiased and its variance is $\sigma^2$.  By the
  Cram\'er-Rao Theorem,
  \begin{equation*}
    \mathrm{var}(\hat{z}) \triangleq \sigma^2 \geq
    \frac{1}{F_z}.
  \end{equation*}
  Therefore, $F_z\geq\frac{1}{\sigma^2}$ with equality being
  achieved when $z \sim \mathcal{N}(0,\sigma^2)$. \qed
\end{lem}
By Lemma~\ref{lem:MinFI}, the maximum CRB (over all distributions of
$f$ with variance $2\sigma_f^2$) is
\begin{equation}
  \label{eq:CRB1}
  C_f = \left(\frac{2\pi^2}{\sigma^2}
    \| \mathbb{P}_{\bf X}^{\perp}{\bf D}_\lpha{\bf Xh}\|^2 +
\frac{1}{2\sigma_f^2} \right)^{-1}.
\end{equation}

\subsection{MAP Estimator of frequency}

As a result of the preceding analysis, we use a Gaussian prior
distribution on $f$ to calculate the MAP estimator. This choice of
prior represents the least informative prior of all distributions with
variance $2\sigma_f^2$ and mean zero. For a particular channel gain
${\bf h}$, the log-likelihood of the data is
\begin{align}
  L({\bf y}, f) &= \ln p({\bf y},f) = \ln p({\bf y}|f)+\ln p(f)
  \nonumber\\ &\propto \frac{-1}{\sigma^2}\|{\bf y} - {\bf V}_f{\bf
    Xh}\|^2 + \frac{1}{4\sigma_f^2}f^2.
  \label{eq:like1}
\end{align}
The apparent additional factor of two associated with $\sigma_f^2$ is due to
the fact that $f$ has a \emph{real} Gaussian distribution as opposed to
complex (as in the first term above).  For any given frequency, the
maximum of this expression over ${\bf h}$ is achieved when
\begin{equation}
  \label{eq:hest1}
  \hat{\bf h}(f) = ({\bf X}^*{\bf X})^{-1}{\bf X}^*{\bf V}^*_f{\bf y}.
\end{equation}
To find the MAP estimator of $f$, we substitute~\eqref{eq:hest1}
into~\eqref{eq:like1} and minimize the negative,
\begin{equation}
  \label{eq:ML1}
  \hat{f} = \arg\min_{f}\left\{ \|
  \mathbb{P}_{\bf X}^{\perp}{\bf V}^*_f{\bf y}\|^2 +
  \frac{\sigma^2}{4\sigma_f^2}f^2 \right\}.
\end{equation}
We note that as $\sigma_f$ goes to infinity (no prior information),
the estimator~\eqref{eq:ML1} is the standard ML frequency
estimator~\cite{ScharfBook}.

\section{Cooperation Phase}
\label{sec:coop}

In the cooperation phase, the destination node receives the
superposition of signals coming from the source and relay.  Each of
these signals is transmitted with a slightly different frequency due
to system imperfections.  The purpose of this section is to derive a
mini-max estimator for the two frequency offsets $f_{\sd}$ and
$f_{\rd}$.  The estimator is mini-max in the sense that we design the
(asymptotically) minimum variance estimator given that the prior
distribution on the frequencies maximizes the estimator variance.  We
show there exists an optimal transmit frequency for the relay, which
reduces the variance of frequency estimation at the destination.

As the relay has an estimate of $f_{\sr}$ (which is correlated with
$f_{\sd}$ and $f_{\rd}$) this information is useful in reducing the
variance of the estimate at the destination.  We assume the frequency
transmitted from the relay is adjusted according to the following
rule,
\begin{align}
  \label{eq:CoopFreq}
  f_{\rno,\text{Tx}} &\triangleq f_\rno - \gamma\hat{f}_{\sr} \nonumber\\
  &= f_\rno - \gamma (f_{\sr} + e_{\sr})
\end{align}
where $\gamma$ is a parameter to be optimized and $e_{\sr} \triangleq
\hat{f}_{\sr} - f_{\sr}$ is the estimation error from the listening
phase.  We choose this rule as it is a linear function of the estimate
and thus analytically tractable.  When $\gamma = 0$, no frequency
adjustment is made (\emph{e.g.,} when the estimate $\hat{f}_{\sr}$
provides no information about the source's frequency), and when
$\gamma = 1$, the relay transmits its own estimate of the source's
frequency (thus trusting the estimate to provide all of the
information available about the source's frequency).  We now express
the frequency difference between the destination and the relay as
\begin{align}
  f_{\rd} &= f_\dno - f_{\rno,\text{Tx}} \nonumber \\
  & =  f_{\sd} - (1-\gamma) f_{\sr} + \gamma e_{\sr} \label{eq:CoopFreq1}. 
\end{align}
The two frequencies to be estimated at the destination node are
$f_{\rd}$ and $f_{\sd}$.

\subsection{Covariance of frequencies}

Before calculating the MAP estimator of $f_{\sd}$ and $f_{\rd}$, we
compute the least informative joint prior distribution.  First, the
covariance matrix of these random variables is found and then we show
that the joint Gaussian distribution is the least informative prior.

To proceed, we calculate the covariance matrix of $f_{\sd}$,
$f_{\sr}$, and $e_{\sr}$. The mean of $f_{\sd}$ and $f_{\sr}$ are
zero, $\mathbb{E}(f_{\sd}^2) = \mathbb{E}(f_{\sr}^2) = 2\sigma_f^2$
and $\mathbb{E}(f_{\sd}f_{\sr}) = \sigma_f^2$.  Now consider
$\mathbb{E}(e_{\sr})$ (we show here that the MAP estimator derived
above is asymptotically unbiased, \emph{i.e.,} $\mathbb{E}(e_{\sr}) =
0$ for high SNR).  Using the definition of $e_{\sr}$
and~\eqref{eq:ML1},
\begin{align}
  e_{\sr} & =  -f_{\sr} + \xi\nonumber\\
  \xi & =  
  \arg\min_{f}\left\{ \|{\bf V}_f
  \mathbb{P}_{{\bf X}_\lpha}^{\perp}{\bf V}_f^*{\bf y}\|^2 +
  \frac{\sigma_\rno^2}{4\sigma_f^2}f^2 \right\}.
  \nonumber
\end{align}
By expressing the expectation as
\begin{equation*}
  \mathbb{E}(e_{\sr}) =
  \mathbb{E}_{f_{\sr}}(\mathbb{E}_{e_{\sr}|f_{\sr}}(\xi -
  f_{\sr}|f_{\sr})),
\end{equation*}
the conditional expectation $\mathbb{E}_{e_{\sr}|f_{\sr}}(\xi|f_{\sr})$
needs to be calculated.  Continuing the asymptotic analysis, for high
SNR, we replace $\bf{y}$ with its mean and obtain
\begin{align}
  \label{eq:CondMean}
  \mathbb{E}_{e_{\sr}|f_{\sr}}(\xi|f_{\sr}) \approx & \arg\min_{f}\Big\{
  \|{\bf V}_f \mathbb{P}_{{\bf X}_\lpha}^{\perp}{\bf V}^*_f{\bf
    V}_{f_{\sr}}{\bf X}_\lpha{\bf h}_{\sr}\|^2\nonumber\\ &  +
  \frac{\sigma_\rno^2}{4\sigma_f^2}f^2 \Big\},
\end{align}
where the approximation is exact in the limit
$\sigma_\rno^2\rightarrow 0$.  We perform the change of variables
$\tilde{f}_{\sr} = 0$ and $\tilde{f} = f-f_{\sr}$, therefore, ${\bf
  V}_{\tilde{f}_{\sr}} = {\bf I}$.  The first term
in~\eqref{eq:CondMean} is
\begin{equation*}
  {\bf h}_{\sr}^*{\bf X}_\lpha^*{\bf V}_{\tilde{f}} \mathbb{P}_{{\bf
    X}_\lpha}^{\perp}{\bf V}^*_{\tilde{f}}{\bf X}_\lpha{\bf h}_{\sr},
\end{equation*}
which is greater than or equal to zero and only equal to zero when
$\tilde{f} = 0$ (\emph{i.e.,} $f = f_{\sr}$).  This function is thus
locally convex about the point $f = f_{\sr}$ and therefore locally
quadratic.  The second order Taylor series approximation is
\begin{equation*}
  \underbrace{\pi^2 \|\mathbb{P}_{{\bf X}_\lpha}^{\perp}{\bf D}_\lpha{\bf
      X}_\lpha {{\bf h}_\sr}\|^2}_{Q} \tilde{f}^2.
\end{equation*}
The value $Q$ can be considered the effective signal power including
all system and estimation gains.  Returning to~\eqref{eq:CondMean},
\begin{align}
  \label{eq:Kdef}
  \mathbb{E}(\xi|f_{\sr}) &\approx \arg\min_f\left\{ Q\cdot(f-f_{\sr})^2 +
  \underbrace{\frac{\sigma_\rno^2}{4\sigma_f^2}}_{K}f^2\right\}\nonumber\\ & = 
  \frac{Q }{Q+K}f_{\sr}.
\end{align}
Completing the mean of $e_{\sr}$,
\begin{equation*}
  \mathbb{E}(e_{\sr}) = \mathbb{E}\left(\frac{Q}{Q+K}f_{\sr} - f_{\sr}\right) = 0
\end{equation*}
because the mean of $f_{\sr}$ is zero and thus the estimator is
asymptotically unbiased.

Continuing on with the covariance,
\begin{equation*}
\mathbb{E}(f_{\sr}e_{\sr}) = \mathbb{E}(f_{\sr}\mathbb{E}(e_{\sr}|f_{\sr}))
= \frac{-2K}{Q+K}\sigma_f^2
\end{equation*}
and similarly $\mathbb{E}(f_{\sd}e_{\sr}) = \frac{-K}{Q+K}\sigma_f^2$
where $K$ is defined in~\eqref{eq:Kdef}.  Following a similar argument
as above for $\mathbb{E}(e_{\sr}^2)$ yields the result that the
variance of $e_{\sr}$ is $\frac{2K}{Q+K}\sigma_f^2$, which is equal to
the CRB in~\eqref{eq:CRB1}.  Thus~\eqref{eq:ML1} is an asymptotically
efficient estimate of the frequency.  In summary,
\begin{equation*}
  \mathrm{Cov}(f_{\sd},f_{\sr},e_{\sr}) = \sigma_f^2\left[
  \begin{array}{ccc}
    2 & 1 & \frac{-K}{Q+K}\\
    1 & 2 & \frac{-2K}{Q+K}\\
    \frac{-K}{Q+K} & \frac{-2K}{Q+K} & \frac{2K}{Q+K}
  \end{array}
  \right].
\end{equation*}

With this covariance matrix calculated, the covariance of $f_{\sd}$ and
$f_{\rd}$ is 
\begin{equation}
\label{eq:FreqCov}
{\bf R}_{f_{\sd},f_{\rd}} \triangleq \sigma_f^2\left[
  \begin{array}{cc}
    2 & \frac{(1+\gamma)Q +K}{Q+K}\\
    \frac{(1+\gamma)Q+K}{Q+K} & 2\frac{(1-\gamma + \gamma^2)Q+K}{Q+K}
  \end{array}
  \right].
\end{equation}

\subsection{Cram\'er-Rao Bound in Cooperative Phase}

Recall the signal models for the cooperation phase~\eqref{eq:CoopSig}
and the listening phase~\eqref{eq:ListenSig} as well as the relation
between the two frequencies to be estimated $f_{\rd}$ and
$f_{\sd}$~\eqref{eq:CoopFreq1}.  The unknown parameters are $f_{\sd}$,
${f_{\rd}}$, ${\bf h}_{\sdc}$, ${\bf h}_{\rd}$, and ${\bf h}_{\sdl}$.
For compactness, define ${\bf f} = [f_\sd \ \; f_\rd]^T$.  The
deterministic FIM (${\bf J}_{\theta|{\bf f}}$) is a $(2+6P) \times
(2+6P)$ matrix with the structure of~\eqref{eq:FIMStructure} where
${\bf \Delta}$ is $2\times 2$.  Given the frequency random variables,
the distributions of ${\bf y}_\cpha$ and ${\bf y}_{\sdl}$ are
independent and the joint distribution is written as
\begin{equation*}
  p({\bf y}_\cpha,{\bf y}_\sdl,{\bf f}) = p({\bf y}_\cpha|{{\bf f}})
  p({\bf y}_\sdl|{{\bf f}})p({{\bf f}})
\end{equation*}
and the FIM is written as
\begin{equation*}
  {\bf J}_{\theta} = {\bf J}_{\theta|{\bf f}}({\bf y}_\cpha) + 
  {\bf J}_{\theta|{\bf f}}({\bf y}_\sdl) + {\bf J}_{{\bf f}}.
\end{equation*}
The blocks of the matrix ${\bf J}_{\theta|{\bf f}}({\bf y}_\cpha)$ are
\begin{align}
  {\bf \Delta}_{11,\cpha}& = \frac{2\pi^2}{\sigma_\dno^2}\|{\bf
    D}_\cpha{\bf X}_\sdc{\bf h}_\sdc\|^2 \nonumber \\
  {\bf \Delta}_{22,\cpha}& = \frac{2\pi^2}{\sigma_\dno^2}\|{\bf
    D}_\cpha{\bf X}_\rd{\bf h}_\rd\|^2 \nonumber \\
  {\bf \Delta}_{12,\cpha} = {\bf \Delta}_{21,\cpha} &=
  \frac{2\pi^2}{\sigma_\dno^2} \Re\left\{{\bf h}_\sdc^*{\bf
    X}_\sdc^*{\bf V}_{f_\sd}^* {\bf V}_{f_\rd}{\bf D}_\cpha^2{\bf
    X}_\rd{\bf h}_\rd\right\}\nonumber\\
  {\bf \Xi}_{11,_\cpha} & =  \frac{1}{\sigma_\dno^2}{\bf
    X}_\sdc^*{\bf X}_\sdc\nonumber\\
  {\bf \Xi}_{22,_\cpha} & =  \frac{1}{\sigma_\dno^2}{\bf
    X}_\rd^*{\bf X}_\rd\nonumber\\
  {\bf \Xi}_{12,_\cpha} = {\bf \Xi}_{21,\cpha}^* & = 
  \frac{1}{\sigma_\dno^2}{\bf X}_\sdc^*{\bf V}_{f_\sd}^*{\bf
    V}_{f_\rd} {\bf X}_\rd \nonumber\\
  {\bf \Lambda}_{11,\cpha} & =  \frac{-j\pi}{\sigma^2_\dno} {\bf
    h}_\sdc^*{\bf X}_\sdc^*{\bf D}_\cpha{\bf X}_\sdc\nonumber\\
  {\bf \Lambda}_{22,\cpha} & =  \frac{-j\pi}{\sigma^2_\dno} {\bf
    h}_\rd^*{\bf X}_\rd^*{\bf D}_\cpha{\bf X}_\rd\nonumber\\
  {\bf \Lambda}_{12,\cpha} & =  \frac{-j\pi}{\sigma^2_\dno} {\bf
    h}_\sdc^*{\bf X}_\sdc^*{\bf V}_{f_\sd}^*{\bf V}_{f_\rd}{\bf D}_\cpha{\bf
    X}_\rd\nonumber\\
  {\bf \Lambda}_{21,\cpha} & =  \frac{-j\pi}{\sigma^2_\dno} {\bf
    h}_\rd^*{\bf X}_\rd^*{\bf V}_{f_\rd}^*{\bf V}_{f_\sd}{\bf D}_\cpha{\bf
    X}_\sdc\nonumber,
\end{align}
and zero for terms not listed.  The diagonal
matrix ${\bf D}_\cpha$ is defined similar to ${\bf D}_\lpha$
in~\eqref{eq:DefD} with $N_\cpha$ replacing $N_\lpha$.

 For data obtained during the listening phase, the matrix ${\bf
   J}_{\theta|{\bf f}}({\bf y}_\sdl)$ is 
\begin{align}
  {\bf \Delta}_{11,\lpha} &= \frac{2\pi^2}{\sigma^2}\|{\bf D}_\lpha
  {\bf X}_\lpha{\bf h}_\sdl\|^2\nonumber\\
  {\bf \Xi}_{33,\lpha}&= \frac{1}{\sigma^2}
  {\bf X}_\lpha^*{\bf X}_\lpha \nonumber\\
  {\bf \Lambda}_{13,\lpha} &= \frac{-j\pi}\sigma^2
    {\bf h}_\sdl^*{\bf X}_\lpha^*{\bf D}_\lpha{\bf X}_\lpha\nonumber
\end{align}
and zero for terms not listed.

To calculate $\mathbb{E}({\bf J}_{\theta|{\bf f}})$, note that only
the $(1,2)$ and $(2,1)$ cross terms of the submatrices above
(\emph{i.e.,} ${\bf \Delta}_{12}$, ${\bf \Xi}_{1,2}$, ${\bf
  \Lambda}_{12}$, \ldots) are dependent on the frequencies.  In each
case, the dependency is of the form ${\bf AV}_{f_\sd}^*{\bf
  V}_{f_\rd}{\bf B}$ where ${\bf A}$ and ${\bf B}$ are deterministic
matrices or vectors.  Looking at the $n^{th}$ term of ${\bf
  V}_{f_\sd}^*{\bf V}_{f_\rd}$,
\begin{equation*}
  \mathbb{E}([{\bf V}_{f_\sd}^*{\bf V}_{f_\rd}]_{nn}) =
  \mathbb{E}(e^{j\pi d_n (f_\rd - f_\sd)})
\end{equation*}
where $d_n = 2n-1-N_\cpha$.  This expectation is just the characteristic
function of the random variable $f_\rd - f_\sd$ evaluated at $\pi d_n$
(denoted as $\Phi_{f_\rd-f_\sd}(\pi d_n)$).  Let $[{\bf M}]_{nn} =
\Phi_{f_\rd-f_\sd}(\pi d_n)$ be a diagonal matrix, then we replace ${\bf
  V}_{f_\sd}^*{\bf V}_{f_\rd}$ with ${\bf M}$ in all cross terms of the FIM
blocks. The FIM is then expressed as
\begin{equation}
  \label{eq:FIM2}
  \text{FIM} = \mathbb{E}({\bf J}_{\theta|{\bf f}}) + {\bf J}_{{\bf f}}
\end{equation}
where ${\bf J}_{{\bf f}}$ is nonzero only in the upper left $2\times
2$ block and this block is equal to ${\bf F}_{{\bf f}}$, the Fisher
information matrix of $f_\sd$ and $f_\rd$.  Using the Shur complement
of the upper left $2\times 2$ block of~\eqref{eq:FIM2}, the CRB for
the frequencies are the diagonal entries of 
\begin{equation}
  \label{eq:CRB2}
  {\bf C}_{{\bf f}} = \left({\bf \Delta -
    2\Re\{{\bf\Lambda\Xi}^{-1}{\bf \Lambda}^*\}} +
  {\bf F}_{{\bf f}}\right)^{-1}.
\end{equation}

In the sequel, we desire to make conclusions about the performance of
the collaborative system based on the derived bounds.  As the absolute
phase of the signal at each node is hard to control and cannot be
relied on to remain stable over time, we find the worst case
CRB and use this in the subsequent discussion.  That is, for ${\bf
  h}_a = \tilde{\bf h}_a e^{j \phi}$, find $\phi$ maximizing the
CRB~\eqref{eq:CRB2}.  The resulting expression is
\begin{equation}
  \label{eq:CRBmax}
  {\bf C}_{{\bf f},\max} = \left(\tilde{\bf \Delta} -
    2\,\text{abs}\{{\bf\Lambda\Xi}^{-1}{\bf \Lambda}^*\} +
  {\bf F}_{{\bf f}}\right)^{-1},
\end{equation}
where $\tilde{\bf \Delta}_{ii} = {\bf \Delta}_{ii}$ and 
\begin{equation}
  \tilde{\bf \Delta}_{12} = \tilde{\bf \Delta}_{21} =
  \frac{-2\pi^2}{\sigma_\dno^2} \text{abs}\left\{{\bf h}_\sdc^*{\bf
    X}_\sdc^*{\bf M}{\bf D}_\cpha^2{\bf
    X}_\rd{\bf h}_\rd\right\}\nonumber.
\end{equation}
Effectively, the phase $\phi$ is chosen to maximize magnitude of the
off-diagonals of the matrix to be inverted in~\eqref{eq:CRBmax}, which
in turn maximizes the diagonals of the inverse (the negative signs are
chosen for the off-diagonal terms because the FIM of the prior
distribution, as calculated in the next section, also has negative
off-diagonal terms).

\subsection{Distribution of Frequencies}

We now desire to find the distribution of $f_\sd$ and $f_\rd$, which
maximizes the CRB for a given frequency covariance ${\bf R}_{{\bf
    f}}$~\eqref{eq:FreqCov}.  In order to do this, we assume the
training sequences are chosen to provide near optimal performance.
Examining~\eqref{eq:CRBmax}, an ideal set of training sequences would
zero out the off-diagonal terms in $\tilde{\bf \Delta}$ and also
zero out the $({\bf\Lambda\Xi}^{-1}{\bf \Lambda}^*)$ term.  Thus for
any constant modulus training sequences, the best CRB is
\begin{equation}
  \label{eq:CRBopt}
  {\bf C}_{{\bf f},\text{opt}} = \left({\bf \Delta}_{\text{opt}} - {\bf
    F}_{\bf f}\right)^{-1}
\end{equation}
where ${\bf \Delta}_\text{opt} = \text{diag}\{{\bf \Delta}\}$.  We
show in Section~\ref{sec:Sim}, by simulation,  sequences exist
where~\eqref{eq:CRBmax} is close to~\eqref{eq:CRBopt}.  Under the
assumption of a good set of sequences, the dependence on the
distribution of $f_\sd$ and $f_\rd$ enters only through ${\bf F}_{{\bf
    f}}$.  We use the following lemma to find the distribution 
maximizing the CRB.
\begin{lem}
  \label{lem:PosDef}
  For ${\bf A}$, ${\bf B}$, and ${\bf C}$ positive definite Hermitian
  matrices, if ${\bf B} >{\bf C}$ (\emph{i.e.,} ${\bf B} - {\bf C}$ is
  positive definite), then $({\bf A} + {\bf C})^{-1} - ({\bf A} + {\bf
    B})^{-1}$ has positive diagonal entries.  \pf By assumption,
  $({\bf A} + {\bf B}) > ({\bf A} + {\bf C})$, which implies 
  $({\bf A} + {\bf C})^{-1} > ({\bf A} + {\bf B})^{-1} $. Thus the
  difference of the matrices is positive definite Hermitian and
  therefore has positive diagonal elements. \qed
\end{lem}
To maximize the diagonal elements of the CRB~\eqref{eq:CRBopt},
Lemma~\ref{lem:PosDef} implies  ${\bf F}_{{\bf f}}$ is as small as
possible.  Using an argument similar to the scalar case of
Lemma~\ref{lem:MinFI}, the Gaussian distribution satisfies this
requirement and ${\bf F}_{{\bf f}} = {\bf R}_{{\bf f}}^{-1}$.  The
assumptions that $f_\sno$, $f_\rno$, $f_\dno$ and the estimation error
from the listening phase $e_{\sr}$ are jointly Gaussian is therefore
the least informative prior given the specified variances and
correlations.

\subsection{Optimal $\gamma$}

With the aim of deriving a mini-max estimator, we desire to choose
$\gamma$ in~\eqref{eq:CoopFreq} to minimize the trace of ${\bf C}_{{\bf
    f}}$~\eqref{eq:CRBopt}.  As this expression is not intuitive,
it is helpful to consider a flat fading model.  For flat fading, $P =
1$ and the terms in the optimal CRB~\eqref{eq:CRBopt} are
\begin{equation*}
  {\bf \Delta}_\text{opt} = \left[\begin{array}{cc} \eta_\cpha S_{\sdc} 
      + \eta_\lpha S_{\sdl} &
      {\bf 0}\\{\bf 0} & \eta_\cpha S_{\rd}
    \end{array}\right]
\end{equation*}
and 
\begin{equation*}
  \label{eq:FreqCovFlat}
  {\bf R}_{{\bf f}} = \left[\begin{array}{cc} 2 &
      \frac{2\eta_\lpha(1+\gamma) S_{\sr}+1/\sigma_f^2}{2\eta_\lpha
        S_{\sr} + 1/\sigma_f^2}\\ \frac{2\eta_\lpha(1+\gamma)
        S_{\sr}+1/\sigma_f^2}{2\eta_\lpha S_{\sr} + 1/\sigma_f^2} &
      2\frac{2\eta_\lpha(1-\gamma+\gamma^2)
        S_{\sr}+1/\sigma_f^2}{2\eta_\lpha S_{\sr} + 1/\sigma_f^2}
    \end{array}
\right]\sigma_f^2.
\end{equation*}
where $\eta_\lpha = \frac{2}{3}\pi^2N_\lpha(N_\lpha^2-1)$ (similarly
for $\eta_\cpha$) and $S_{\rd} = \frac{|h_{\rd}|^2}{\sigma_\dno^2}$ is
the signal to noise ratio of the source-relay link (similarly for
$S_\sdc$, $S_\sdl$, and $S_\sr$).

An exact calculation of the optimal $\gamma$ leads to a long,
complicated expression that depends on the SNR of each link and the
variance of the frequency oscillators.  The expression is omitted here
as it gives no insight into the problem.  Later, we show there is
minimal loss when $\gamma$ is always set to 1.  To gain some insight
into the behavior of $\gamma$, consider two limiting cases for
$\gamma_{opt}$: $\sigma_f\rightarrow\infty$ and $\sigma_f\rightarrow
0$.

\subsubsection{Large $\sigma_f$, or no prior information}

By taking the limit of the expression for $\gamma_{opt}$ as
$\sigma_f\rightarrow\infty$, it can be shown that
$\gamma_{opt}\rightarrow 1$. In this case, the relay transmits at a
frequency equal to its estimate of the source frequency. By choosing
this transmit frequency, the operating frequency of the relay $f_\rno$
is removed from the estimation procedure as it contains no information
about the source-destination frequency.

\subsubsection{Small $\sigma_f$}

When $\sigma_f\rightarrow 0$ (or when $1/\sigma_f^2$ is much larger
than any of the link SNRs perhaps due to poor channel SNRs), the CRB
is minimized when $\gamma = 1/2$.  By looking at the MAP frequency
estimator~\eqref{eq:ML1} for this limiting case, the frequency
estimate is zero.  Therefore, no matter what $\gamma$ is chosen, the
relay just transmits at its own frequency.  When $\sigma_f$ is small
(but not zero), there is still some information in the frequency
estimate about the source frequency (besides the information from the
local oscillator model), and by choosing $\gamma \approx 1/2$, both
sources of information are used to select the best transmit frequency.

\begin{figure}
  \centering
  \includegraphics[width=3.2in]{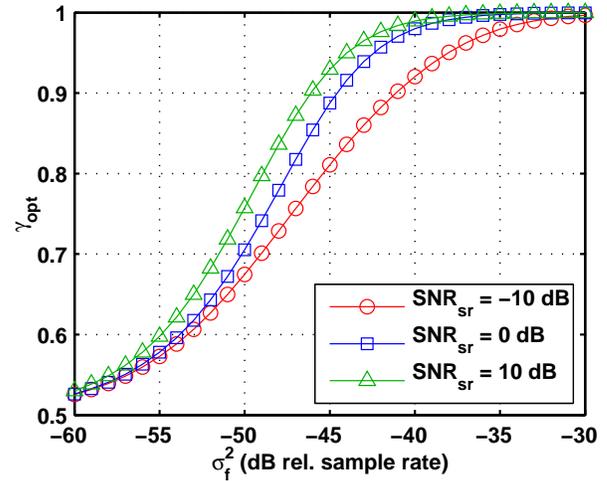}
  \caption{Plot of optimal $\gamma$ as a function of modeled frequency
    variation.  Three curves are shown for different values of gain
    between the source and relay.  The SNR of the source-destination
    and relay-destination node pairs are held constant at 0 dB.}
  \label{fig:gamoptRS}
\end{figure}

As as example of the function $\gamma_{opt}$,
Figure~\ref{fig:gamoptRS} shows plots of several curves of
$\gamma_{opt}$ versus $\sigma_f^2$.
The length of the training signal is $N_\lpha = N_\cpha = 16$ and the
SNR of the source-destination link is $-3$~dB (combining the listening
and cooperation phases, the effective SNR is $0$~dB).  The SNR from
relay to destination, $S_{\rd}$, is also $0$~dB and there is one curve
each for $S_{\sr} \in \{-10 \ \mathrm{dB}, \ 0 \ \mathrm{dB}, \ 10
\ \mathrm{dB}\}$.  For each curve, the transition from $\gamma_{opt} =
1$ to $\gamma_{opt} = 1/2$ appears to occur roughly when $\sigma_f^2
\approx (\eta_\cpha S_{\sdc} + \eta_\lpha S_{\sdl})$ or $\sigma_f^2
\approx \eta_\cpha S_{\rd}$.  These values of $\sigma_f^2$ are
significant because, for example, when $\sigma_f^2 < \eta_\lpha
S_{\sdl}$ (left half of the plot), the assumed prior knowledge of
frequency has more weight than the data, whereas when $\sigma_f^2 >
\eta_\lpha S_{\sdl}$, the information in the data is more important
than the prior model.

\subsection{MAP Estimator of $f_{\sd}$ and $f_{\rd}$}

To calculate the MAP estimate of $f_{\sd}$ and $f_{\rd}$ at the
destination node during the cooperation phase, the covariance between
these two random variables~\eqref{eq:FreqCov} is needed.  Therefore,
the values of $Q$ and $K$ need to be forwarded to the destination
node.  The log-likelihood of the data at the destination is
\begin{align}
  \label{eq:LikeCoop}
  L({\bf y}_\cpha,{\bf y}_\sdl,{\bf f}) =& \ln p({\bf y}_\cpha|{\bf
    f}) + \ln p({\bf y}_\sdl|{\bf f}) + p({\bf f})
  \nonumber\\ \propto& \frac{-1}{\sigma_\dno^2}\|{\bf y}_\sdl - {\bf
    V}_{f_\sd}{\bf X}_\lpha{\bf h}_\sdl\|^2 \\ & -
  \frac{1}{\sigma_\dno^2}\left\|{\bf y}_\cpha - {\mathbb X}({\bf
    f}){\bf g} \right\|^2 - \frac{1}{2}{\bf f}^T{\bf R}^{-1}_{\bf
    f}{\bf f},\nonumber
\end{align}
where ${\mathbb X}({\bf f}) \triangleq \left[\begin{array}{cc} {\bf
      V}_{f_\sd}{\bf X}_\sdc & {\bf V}_{f_\rd}{\bf X}_\rd
    \end{array}\right]$ and ${\bf g}^T = 
\left[ \begin{array}{cc} {\bf h}_\sdc^T &{\bf
      h}_\rd^T \end{array}\right]$.  As before, choose estimates for
${\bf g}$ and ${\bf h}_\sdl$ to maximize the likelihood for any given
frequency pair,
\begin{align*}
  \hat{\bf g}({\bf f}) &= 
  (\mathbb{X}^*\mathbb{X})^{-1}\mathbb{X}^*{\bf y}_\cpha \\
  \hat{\bf h}_\sdl(f_\sd) &= 
  ({\bf X}_\lpha^*{\bf X}_\lpha)^{-1}
  {\bf X}_\lpha^*{\bf V}_{f_\sd}^*{\bf y}_\sdl.
\end{align*}
Substituting these estimates into~\eqref{eq:LikeCoop} and minimizing the
negative to obtain the MAP frequency estimator
\begin{align}
  \label{eq:MAP2}
  \hat{\bf f} =& \arg\min_{{\bf f}}\Big\{ \|
  \mathbb{P}_{{\mathbb X}({\bf f})}^{\perp}{\bf y}_\cpha\|^2 +
  \|\mathbb{P}_{{\bf X}_\lpha}^{\perp}{\bf V}_{f_\sd}{\bf y}_\sdl\|^2 +
  \nonumber \\
   & + \ \frac{\sigma_\dno^2}{2}{\bf f}^T{\bf R}^{-1}_{\bf f}{\bf f} \Big\}.
\end{align}

We note the special case of $\sigma_f\rightarrow\infty$ (which
implies $\gamma_{opt} = 1$).  For $\gamma = 1$, the
covariance~\eqref{eq:FreqCov} needed in the MAP estimator simplifies
to
\begin{equation*}
{\bf R}_{f_{\sd},f_{\rd}}
 = \sigma_f^2\left[
  \begin{array}{cc}
    2 & \frac{2Q +K}{Q+K}\\
    \frac{2Q+K}{Q+K} & 2
  \end{array}
  \right],
\end{equation*}
which has a finite inverse when $\sigma_f<\infty$.  However, when
$\sigma_f \rightarrow \infty$, we evaluate the limit of ${\bf R}_{\bf
  f}^{-1}$ resulting in
\begin{equation*}
\lim_{\sigma_f\rightarrow \infty}{\bf R}_{\bf f} = {\bf
  \zeta\zeta}^T\frac{2 Q}{\sigma_\rno^2} = {\bf
  \zeta\zeta}^T\underbrace{\frac{2\pi^2}{\sigma_\rno^2}\|\mathbb{P}_{{\bf
      X}_\lpha}^{\perp}{\bf D}_\lpha{\bf X}_\lpha{\bf h}_{\sr}\|^2}_{C_{f_\sr}}
\end{equation*}
where $\zeta^T = [1 \ -1]$ and $C_{f_\sr}$ is the CRB of the frequency in
the source-relay link~\eqref{eq:CRB1} with $\sigma_f = \infty$.
The penalty term (last term) of the MAP estimator~\eqref{eq:MAP2}
simplifies to
\begin{equation*}
\frac{\sigma_\dno^2}{2}{\bf f}^T{\bf R}^{-1}_{\bf f}{\bf f}\rightarrow
\frac{\sigma_\dno^2}{2C_{f_\sr}}(f_\sd - f_\rd)^2.
\end{equation*}
Thus the penalty term is a quadratic of the frequency difference term
normalized by the ratio of error variances (noise power over frequency
estimation error variance).

\section{Simulations}
\label{sec:Sim}

In the previous section, we showed the optimal $\gamma$ for extreme
values of $\sigma_f$ is either $1$ or $1/2$ and when $\gamma_{opt}$
approaches $1/2$, its effect is small because the frequency adjustment
is going toward zero.  In this section, we show by simulation, the
penalty for choosing $\gamma = 1$ instead of $\gamma = \gamma_{opt}$
is usually limited to a few tenths of a decibel.  Thus, near optimal
performance is achieved without communicating any of the link SNRs
back to the relay for calculation of $\gamma_{opt}$.  We also show the
existence of training sequences where~\eqref{eq:CRBmax} is close
to~\eqref{eq:CRBopt}.  Finally, we show the benefit of letting the
relay set its transmit frequency based on information received during
the listening phase.

We simulate a three node system in a frequency flat environment.  In
all simulations, we use the SNR of the $\sd$ link (assuming $S_{\sdc}
= S_\sdl$) as a reference value.  The following configuration is
considered: let $S_\rd = S_\sdc$ and then vary the link SNR of the
source-relay link relative to $S_\sdc$.  Let $N_\lpha = N_\cpha$.  The
prior distribution for the operating frequency we assume is Gaussian
with a variance of $-40$~dB relative to the sample rate (\emph{e.g.,}
a 2 parts-per-million variance of a local oscillator at 900 MHz with
4.5 MHz sample rate~\cite{ShinCKT05}).

For flat fading channels and constant modulus training sequences, it
is sufficient to choose ${\bf x}_\lpha = {\bf 1}$ (the vector of all
ones) and ${\bf x}_\sdc = {\bf 1}$.  A search is performed to find
${\bf x}_\rd$ which minimizes the CRB~\eqref{eq:CRBmax}.  For values of
$N_\cpha \in \{4,\ 8,\ 16\}$ an exhaustive search over all binary
sequences is performed (results hold independent of choice between
$\gamma = \gamma_{opt}$ or $\gamma = 1$) and for values of
$N_\cpha>16$, a randomized search over binary sequences is performed.
For each value of $N_\cpha$ (up to 128) the optimal sequence
for ${\bf x}_\rd$ has the following structure:
\subsubsection*{Sequence Design}
Let ${\bf a}_1 = [1, \ -1]^T$ and 
\begin{equation*}
{\bf a}_n^T = [{\bf a}_{n-1}^T, \ -{\bf a}_{n-1}^T]
\end{equation*}
 where ${\bf a}_n$ is length $2^n$ and is the
last column of a Sylvester matrix.  Then the length $N_\cpha = 2^n$
optimal sequence is
\begin{equation*}
{\bf x}_{\rd,opt} = \left[\begin{array}{c}
    {\bf a}_{n-1} \\ -{\bf Ja}_{n-1}
  \end{array}
  \right]
\end{equation*}
where ${\bf J}$ is the exchange matrix which reverses the order of
elements in the vector it multiplies.  

\begin{figure}
  \centering
  \includegraphics[width=3.2in]{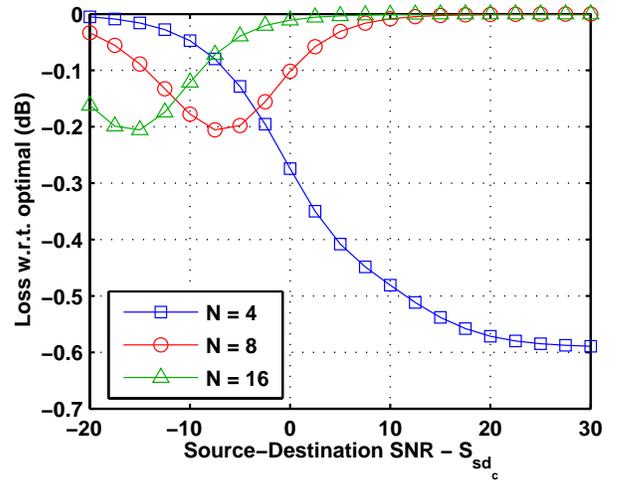}
  \caption{Plot of the loss in performance caused by binary training
    sequence as opposed to an arbitrary sequence, and when choosing
    $\gamma = 1$ versus $\gamma = \gamma_{opt}$.  Relay-destination
    and source-destination SNRs are the same and source-relay SNR is
    10 dB higher.}
  \label{fig:CRBSimRSDiff}
\end{figure}

For the configuration described above, and with $S_\sr = 10\, S_\sdc$,
Figure~\ref{fig:CRBSimRSDiff} shows the difference between the best
possible CRB~\eqref{eq:CRBopt} (for any constant modulus sequence and
$\gamma = \gamma_{opt}$) and the worst case CRB~\eqref{eq:CRBmax} using
the binary sequence shown above and $\gamma = 1$.  The $0.6$~dB
difference for $N_\cpha = 4$ is primarily due to a non-optimal
sequence ${\bf x}_\rd$, whereas the $0.2$~dB difference for other
values of $N_\cpha$ is due to choosing $\gamma = 1$ instead of the
optimal value.  The loss in performance due to a non-optimal sequence
decreases dramatically as $N_\cpha$ increases.  These loss values are
typical of other system configurations as well.  The system behavior
as a function of training sequence illustrates the fact that the CRB
is insensitive to the selection of these sequences.

\begin{figure}
  \centering
  \includegraphics[width=3.2in]{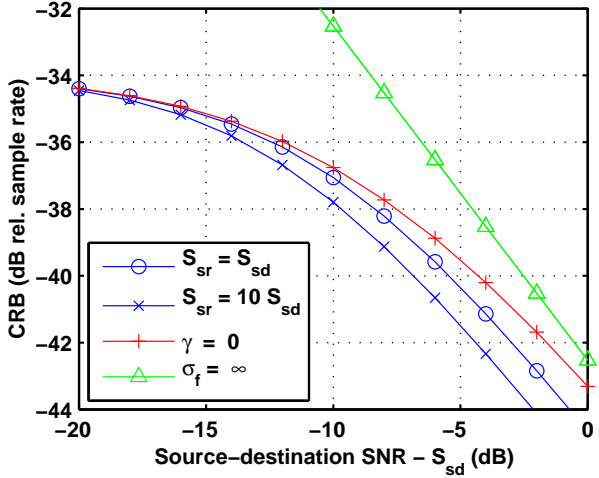}
  \caption{Plot of the sum of Cram\'er-Rao Bounds for $f_{\sd}$ and
    $f_{\rd}$.  Circle and ``x''-marks show bound when $\sigma_f^2 =
    -40$~dB and $\gamma = \gamma_{opt}$, plus marks show bound when
    $\gamma = 0$, and triangles show bound when $\gamma = 0$ and
    $\sigma_f = \infty$ (the standard frequency bound assuming no
    prior information).  All curves are for a length 16 training
    sequence.}
  \label{fig:CRBSimRS}
\end{figure}

Figure~\ref{fig:CRBSimRS} shows the sum of the CRB for the two
frequencies estimated at the destination node as a function of
$S_{\sdc}$.  For this figure, the SNRs of the source-destination link
($S_{\sd}$) and relay-destination link ($S_{\rd}$) are the same.  The
circle and ``x''-marks show the CRB when the SNR of the source-relay
link $S_{\sr}$ is, respectively, the same as and 10~dB higher than
$S_{\sd}$.  The plus marks show the CRB when $\gamma = 0$.  The
difference between the plus-marks and the circle and ``x''-marks show
the potential gain in estimation performance by changing the relay's
transmit frequency (greater benefit when the SNR is large).  The
triangles show the CRB when no prior information is used.  This
shows a great advantage of using a prior model when the SNR is low.

\section{Sub-Optimal Algorithms}
\label{sec:Algs}

The maximum-likelihood frequency estimator~\eqref{eq:MAP2} requires a
two-dimensional search over the frequency range of interest.  As this
is a computationally expensive approach to estimation, we compare the
mean squared error~(MSE) performance of more efficient, sub-optimal
estimation algorithms and introduce a correlation based estimator as
the best compromise between estimation performance and computational
efficiency.  In the remainder of this section, we describe the use of
the one-dimensional ML algorithm as applied to the two signal case and
the correlation algorithm for frequency estimation and compare their
performance.

\subsection{One-Dimensional ML}

As a result of the choice of the training sequence, estimation of the two
frequencies is nearly uncoupled.  Therefore, performing two
independent one-dimensional ML searches for the frequencies is
approximately the same as performing the full two-dimensional ML
search as required by the ML algorithm.  Given the data vector ${\bf
  y}_\cpha$, the one-dimensional ML estimates of the frequencies are
\begin{align}
  \label{eq:ML1frd}
  \tilde{f}_\rd =& \arg\min_f\left\{\|\mathbb{P}_{{\bf
      x}_\rd}^\perp{\bf V}_f^*{\bf y}_\cpha\|^2 +
  \frac{\sigma_\dno}{4\sigma_f}f^2\right\} \\
  \tilde{f}_\sd =& \arg\min_f\{\|\mathbb{P}_{{\bf x}_\lpha}^\perp{\bf
    V}_f^*{\bf y}_\sdl\|^2 + \nonumber\\
   & +\ \|\mathbb{P}_{{\bf x}_\sdc}^\perp{\bf V}_f^*{\bf y}_\cpha\|^2
  + \frac{\sigma_\dno}{4\sigma_f}f^2\},
  \label{eq:ML1fsd}
\end{align}
which do not take the correlations between the frequencies into
account.  To improve the estimates~\eqref{eq:ML1frd}
and~\eqref{eq:ML1fsd}, we assume the variance of each estimate meets
the CRB assuming the prior information is uncorrelated for each
frequency:
\begin{equation*}
  \tilde{\bf C}_{{\bf f}} = \left({\bf \Delta -
    2\Re\{{\bf\Lambda\Xi}^{-1}{\bf \Lambda}^*\}} + \mathrm{diag}\{{\bf
    R}_{\bf f}\}^{-1}\right)^{-1},
\end{equation*}
where $\mathrm{diag}\{{\bf R_f}\}$ is a diagonal matrix consisting of
the diagonal entries of ${\bf R_f}$ (zeroing out the other elements).
This assumption is valid for high SNR and large $N_\cpha$.
Incorporating this knowledge with the prior information, the least
squares estimates of the frequencies are
\begin{equation}
  \label{eq:FinalML1}
  \left[\begin{array}{c}
      \hat{f}_{\sd,ML1}\\\hat{f}_{\rd,ML1} \end{array} \right] =
  {\bf R}_{\bf f}({\bf R}_{\bf f} + \tilde{\bf C}_{\bf f} )^{-1}
    \left[\begin{array}{c}
      \tilde{f}_\sd\\\tilde{f}_\rd \end{array} \right].
\end{equation}

\subsection{Correlation Method}

We first describe a standard correlation frequency estimation method
as presented in~\cite{LuiseR95} and then provide an extension to allow
this algorithm to work in the presence of two signals with known
training sequences.  Assuming a single signal in the presence of flat
fading
\begin{equation*}
  y[n] = e^{j2\pi f n}x[n] + w[n], \quad 1\leq n \leq N.
\end{equation*}
The estimated autocorrelation sequence of $y[n]$ is 
\begin{equation*}
  R[k] = \frac{1}{N-k}\sum_{i = k+1}^N (y[n]\bar{x}[n])(\bar{y}[i-k]x[i-k]).
\end{equation*}
The estimate of the frequency is calculated as
\begin{equation}
  \label{eq:CorrAlg}
  \hat{f} = \frac{1}{\pi(M+1)}\arg\left\{\sum_{k=1}^M R[k] \right\}
\end{equation}
where $M$ is a design parameter and the frequency estimate is
unambiguous if
\begin{equation*}
  |f|<\frac{1}{M+1}.
\end{equation*}
Therefore, $M$ trades performance for estimation range.  The
performance of this algorithm~\eqref{eq:CorrAlg} is shown
in~\cite{LuiseR95} to be close to the CRB when $M = N/2$.  To ensure
adequate estimation range, the maximum allowed value of $M$ is 12
(corresponding to a range of five standard deviations away from the
mean of the prior).  To incorporate the known prior knowledge of the
frequency variance, the estimate~\eqref{eq:CorrAlg} is adjusted
according to the following rule
\begin{equation*}
\hat{f}_p = \frac{2\sigma_f^2}{2\sigma_f^2 + c_f^2}\hat{f}
\end{equation*}
where $c_f^2$ is the CRB of the frequency estimate with no prior
information.  Let
\begin{equation*}
  \hat{f} = \rho({\bf y, x},\sigma_f)
\end{equation*}
be a function that inputs the data vector ${\bf y}$, training vector
${\bf x}$, and prior information, and outputs the frequency estimate
according to the above algorithm.  This algorithm is used without
modification during the listening phase to calculate the estimate
$\hat{f}_\sr = \rho({\bf y}_\sr,{\bf x}_\lpha,\sigma_f)$.

\begin{figure}
  \centering
  \includegraphics[width=3.2in]{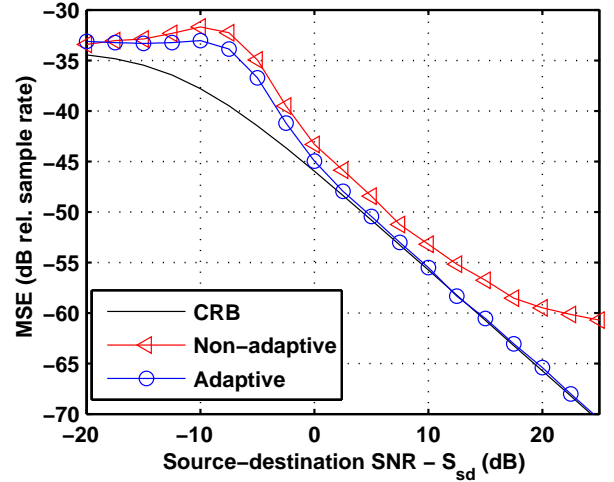}
  \caption{Plot of mean squared error of non-adaptive (circles) and
    adaptive two-step (triangles) correlation algorithms.  The mean
    squared error is compared with the CRB.}
  \label{fig:AdaptCorr}
\end{figure}

For the cooperation phase, there are two signals present and the
undesired signal acts as interference for the desired signal being
estimated.  The estimates provided by the correlation algorithm are
\begin{align}
  \label{eq:HTEsd1}
  \tilde{f}_{\sd,1} =& \rho({\bf y}_\cpha,{\bf x}_\sd,\sigma_f) \\
  \label{eq:HTErd1}
  \tilde{f}_{\rd,1} =& \rho({\bf y}_\cpha,{\bf x}_\rd,\sigma_f),
\end{align}
which exhibit a floor in MSE (see Figure~\ref{fig:AdaptCorr}).  To
improve the estimates, we project out the undesired signal in the
following manner:
\begin{align*}
  \tilde{\bf y}_{\cpha,\sd} =& \mathbb{P}_{{\bf
      V}_{\tilde{f}_{\rd,1}}{\bf x}_\rd }^\perp {\bf y}_\cpha\\
\tilde{\bf y}_{\cpha,\rd} =& \mathbb{P}_{{\bf
    V}_{\tilde{f}_{\sd,1}}{\bf x}_\sd }^\perp {\bf y}_\cpha,
\end{align*}
where the frequency estimates in~\eqref{eq:HTEsd1}
and~\eqref{eq:HTErd1} are used to calculate the interference signal,
which is projected out.  The correlation algorithm is run a second
time to find
\begin{align*}
  \tilde{f}_{\sd,2} =& \rho(\tilde{\bf y}_{\cpha,\sd},{\bf x}_\sd,\sigma_f) \\
  \tilde{f}_{\rd,2} =& \rho(\tilde{\bf y}_{\cpha,\rd},{\bf x}_\rd,\sigma_f).
\end{align*}
The final frequency estimates, with all prior information accounted
for, is calculated similarly to~\eqref{eq:FinalML1},
\begin{equation*}
  \left[\begin{array}{c}
      \hat{f}_{\sd,corr}\\\hat{f}_{\rd,corr} \end{array} \right] =
  {\bf R}_{\bf f}({\bf R}_{\bf f} + \tilde{\bf C}_{\bf f} )^{-1}
    \left[\begin{array}{c}
      \tilde{f}_{\sd,2}\\\tilde{f}_{\rd,2} \end{array} \right].
\end{equation*}

Figure~\ref{fig:AdaptCorr} shows the total MSE
(summation of errors from $\hat{f}_\sd$ and $\hat{f}_\rd$) of the
correlation algorithm compared with the CRB for $N_\cpha = 16$.  The
triangle markers denote the performance of the algorithm without any
adaptation while the circle markers denote the performance of the
adaptive two-step algorithm described above.  For lower SNRs, the
adaptive algorithm has about a 3~dB advantage while the performance
difference is much greater at higher SNRs (above 15~dB).  The
performance of the adaptive algorithm is near optimal.  The slight
``bump'' in performance of the two algorithms at $S_\sd = -10$~dB SNR
is caused by the interaction of the threshold region (the region where
the MSE performance breaks away from the CRB) and the region dominated
by prior information (where the algorithms converge to a $-34$~dB MSE
relative to the sample rate).

\begin{figure}
  \centering
  \includegraphics[width=3.2in]{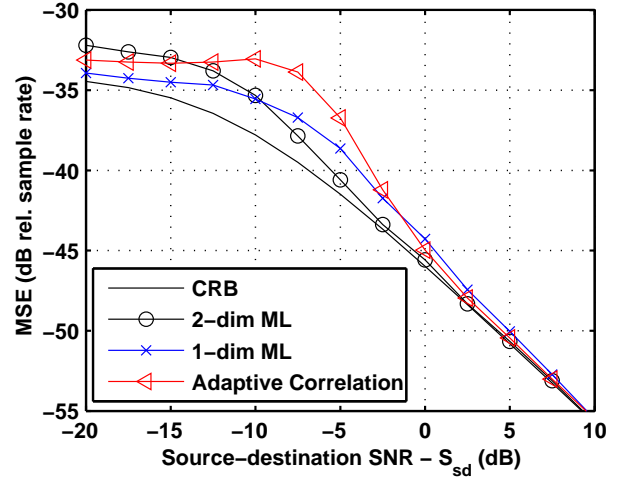}
  \caption{Plot of mean squared error of full (two-dimensional search)
    ML (circles), one-dimensional ML (``x''-marks), and adaptive
    correlation (triangles).  The mean squared error is compared with
    the CRB. }
  \label{fig:AlgComp}
\end{figure}

For the same scenario, Figure~\ref{fig:AlgComp} compares the three
estimation algorithms: full (two-dimensional search) maximum
likelihood (circles), one-dimensional ML (``x''-marks), and the
adaptive correlation algorithm (triangles).  Each of these algorithms
approaches the CRB asymptotically in SNR.  The differences in behavior
at lower SNRs is attributed to the different algorithms entering their
threshold regions at different SNRs.  A more detailed analysis of this
region can be carried out using the methods of~\cite{Richmond06}.

\section{Conclusions}

In this paper, we have derived the Cram\'er-Rao bounds for frequency
offset estimation in a three-node collaborative communication system.
We have shown through simulation, the performance increase obtained by
allowing the relay to change its transmitting frequency.  We have also
shown there exists an optimal transmit frequency for the relay node
based on the other link SNRs and the assumed prior knowledge of the
frequency offsets.  However, there is only a small (tenths of
decibels) penalty if the relay always transmits at its estimate of the
source frequency.  Simulation results also demonstrate the existence
of binary training sequences that result in very little loss as
compared with an arbitrary constant modulus sequence.  We also derived
a computationally efficient correlation based estimation algorithm
that has mean squared error performance close to the CRB.

\bibliography{Bibliog}
\end{document}